\documentclass{elsarticle}
\usepackage{amssymb}
\usepackage{hyperref}

\newcommand{\cmt}[1]{}
\newcommand{\Rs}{R_{sub}}
\newcommand{\Rt}{R_{tot}}

\newcommand{\Rtt}{\tilde{R}_{tot}}

\newcommand{\de}{\partial}

\newcommand{\di}{{\rm d}}
\newcommand{\prot}{\rm{H}}
\newcommand{\he}{\rm{He}}
\newcommand{\cno}{\rm{CNO}}
\newcommand{\fe}{\rm{Fe}}
\newcommand{\degK}{\,\rm{ K}}
\newcommand{\simgt}%
       {\,\hbox{\lower0.6ex\hbox{$\sim$}\llap{\raise0.6ex\hbox{$>$}}}\,}

   \newcommand{\simlt}%
       {\,\hbox{\lower0.6ex\hbox{$\sim$}\llap{\raise0.6ex\hbox{$<$}}}\,}

\newcommand{\be}{\begin{equation}}
\newcommand{\ee}{\end{equation}}

\begin{document}
\begin{frontmatter}

\title{Non-linear diffusive acceleration of heavy nuclei in supernova remnant shocks}

\author{D. Caprioli}
\ead{caprioli@arcetri.astro.it}

\author{P. Blasi}
\ead{blasi@arcetri.astro.it}

\author{E. Amato}
\ead{amato@arcetri.astro.it}

\address{INAF/Osservatorio Astrofisico di Arcetri, Largo E. Fermi, 5 - 50125 Firenze, Italy}

\begin{abstract}
We describe a semi-analytical approach to non-linear diffusive shock acceleration in the case in which nuclei other than protons are also accelerated. The structure of the shock is determined by the complex interplay of all nuclei, and in turn this shock structure determines the spectra of all components. The magnetic field amplification upstream is described as due to streaming instability of all nuclear species. The amplified magnetic field is then taken into account for its dynamical feedback on the shock structure as well as in terms of the induced modification of the velocity of the scattering centers that enters the particle transport equation. The spectra of accelerated particles are steep enough to be compared with observed cosmic ray spectra only if the magnetic field is sufficiently amplified and the scattering centers have high speed in the frame of the background plasma. We discuss the implications of this generalized approach on the structure of the knee in the all-particle cosmic ray spectrum, which we interpret as due to an increasingly heavier chemical composition above $10^{15}$eV. The effects of a non trivial chemical composition at the sources on the gamma ray emission from a supernova remnant when gamma rays are of hadronic origin are also discussed.
\end{abstract}

\begin{keyword}
supernova remnants; shock waves; Galactic cosmic rays; nuclei; knee
\end{keyword}

\end{frontmatter}

\section{Introduction}

A satisfactory understanding of the origin of cosmic rays (CRs) must deal with the issue of chemical composition. Probably the most striking instance of the role played by the different chemicals is the origin of the knee in the all-particle CR spectrum. In a scenario in which the maximum energy of accelerated particles scales with the charge of the particles involved, a knee arises naturally as a superposition of spectra of chemicals with different charges $Ze$. Even more important, the change of spectral slope on the two sides of the knee is determined by the relative abundance of different chemicals as a function of $Z$, convolved with the effects of rigidity dependent propagation in the Galaxy.

While there has been much work on the propagation of nuclei in the Galaxy (mainly because of the importance it has for the prediction of travel time of CRs in the Galaxy and ratios of secondary to primary fluxes), not much attention has been devoted to the acceleration of nuclei in the sources. CR acceleration in SNRs is believed to take place through the mechanism of diffusive shock acceleration, in its non-linear version that allows one to take into account the reaction of accelerated particles on the plasma and on local magnetic fields. Several versions of the non-linear theory of diffusive acceleration at shocks have been developed (see \cite{maldrury} for a review), but most of them include only protons as accelerated particles. Two noticeable exceptions are represented by the work of \cite{berevolk} and that of \cite{pzs10}. In both papers the calculations consist of a numerical solution of the coupled equations of CR transport and conservation of  mass, momentum and energy flux of the overall CR plus background plasma. In the paper by Berezhko and V\"{o}lk \cite{berevolk} the calculations were not illustrated in detail, and it is difficult for us to appreciate the assumptions that were adopted there. The spectra from individual sources was found to be very flat (flatter than $E^{-2}$ at high energies), so that the observed CR spectrum could be recovered only by assuming a Galactic diffusion coefficient as steep as $D(E)\propto E^{0.75}$, which however is not consistent with measurements of the CR anisotropy at the Earth. In the paper by Ptuskin et al. \cite{pzs10}, some more details were provided, and the authors discussed the important role of the velocity of the scattering centers on the shape of the spectrum of accelerated particles, which, as a consequence, is here much steeper than that found by Berezhko and V\"{o}lk \cite{berevolk}.
 
A fully non-linear theory which includes nuclei is made rather complex by at least two issues: (1) nuclei change the structure of the shock, making the problem harder to tackle and (2) the injection of nuclei in the accelerator is more challenging to be modelled than it is for protons, especially because nuclei can be produced as a result of dust sputtering. In a non-linear theory, the second point clearly feeds back onto the first one.

It is important to recall that the injection and the non-linear acceleration of nuclei in the Earth's Bow Shock (EBS) has been successfully described using a Monte Carlo simulation in the pioneering paper in Ref. \cite{emp90}. These calculations showed how the case of the EBS is and remains, 20 years later, the most clear cut instance of occurrence of Non-Linear Diffusive Shock Acceleration (NLDSA) in collisionless shocks. However some caution should be adopted in extending these results to the case of SNRs, in that the physical conditions at the EBS might be somewhat different: the lack of electron injection, the fact that magnetic field amplification in the EBS appears to remain in the quasi-linear regime and the crucial role played by dust sputtering for ion injection in SNRs might be considered as possible evidences of such differences.

The last point is particularly relevant to our problem in that, in order to account for the observed discrepancy between the chemical composition of typical interstellar medium and CRs in our Galaxy, refractory elements (such as Mg, Al, Si and Fe) have to be injected in a preferential way in the acceleration process with respect to volatile elements \cite{dust1, dust3}. 
Refractory elements are usually trapped in dust grains, and their preferential injection has been interpreted as a consequence of the sputtering of the grains when they are swept up by the SNR shock. In particular, the processes which lead to the injection of suprathermal iron nuclei as a consequence of the sputtering of accelerated dust grains have been put forward quantitatively for the first time in \cite{dust2}.   

Nevertheless, a self-consistent description of this process would require both a detailed knowledge of the dust chemistry close to a SNR shock and an accurate physical description of the grain sputtering, along with a time-dependent treatment of the ionization of dust and atoms during their acceleration.
The intrinsic complexity of such a phenomenon led us (as well as the authors of the previous work mentioned above \cite[]{berevolk,pzs10}) to a simplified treatment of the nuclei injection: the measured spectra of relativistic particles at Earth are reproduced without explicitly taking into account the microphysics either of the nuclei ionization or of the dust sputtering (see \S\ref{sec:abunda}).

One might argue that the relatively low abundances of nuclei in the cosmic radiation observed at Earth might make their influence on the shock structure negligible, so that in describing the acceleration process, the shock structure could be treated as determined by protons alone, with nuclei behaving as test particles. However, after correcting for propagation effects, it is easy to show that the nuclear contribution to the total pressure and magnetic field amplification in the vicinity of a typical supernova remnant shock may be as important as that of protons. 

In this paper we describe the generalization of the non-linear theory of DSA developed by Amato and Blasi \cite{amato1,amato2} and Caprioli et al. \cite{bound} to include nuclei of different charges. We calculate the spectrum of all species as accelerated at the shock and the structure of the shock (including magnetic effects) induced by them. We also comment on the implications of acceleration of nuclei on the spectra of secondary products of particle interactions, especially gamma rays. Finally we show the all-particle CR spectrum at Earth resulting from this calculation. 

The calculations discussed here are semi-analytical, and from the computational point of view very inexpensive. This allows us to explore a wide region in parameter space, which is particularly important when dealing with the goal of explaining the CR spectra and chemical abundances observed at Earth. For simplicity we consider here only supernovae exploding in a homogeneous interstellar medium (ISM). While more complex situations can be treated in the context of our formalism, they introduce a wide range of new and hardly accessible parameters, which overshadow the main physical results. We will comment further on this point whenever we deem it necessary.

The paper is organized as follows: in \S~\ref{sec:model} we illustrate the generalization of the equations and solution techniques following the non-linear theory of \cite{amato1,amato2,bound}. In \S~\ref{sec:abunda} we comment on the abundance of nuclei in CRs and at the sources, and how it relates to the injection of the nuclear component at the shock. In \S~\ref{sec:spectra} we illustrate our results for the spectra of accelerated particles and the structure of the shock. In \S~\ref{sec:gamma} we discuss the implications of the presence of accelerated nuclei at the shock for the prediction of a gamma ray flux, as due to production and decay of neutral pions. In \S~\ref{sec:allpart} we compare our findings with the CR flux detected at Earth, discussing the spectra of individual chemicals and the appearance of the knee at $E\sim 10^{6}$ GeV. We discuss our general results and we compare them with previous findings in \S~\ref{sec:discuss}.

\section{Equations and Solution techniques}
\label{sec:model}

In this section we generalize the semi-analytical formalism developed in \cite{amato1,amato2,bound} to the case in which nuclei heavier than Hydrogen (hereafter simply Heavy Nuclei, HN) are also injected and accelerated at a stationary, plane, parallel (background magnetic field parallel to the shock normal), newtonian shock wave.
We label with a subscript $i$ quantities referring to different chemical elements, so that the convection-diffusion equation for the isotropic part of the distribution function, $f_{i}(x,p)$, reads, for each species \citep[see e.g.\ ][]{skillinga}:
\begin{equation}
\tilde{u}(x)\frac{\de f_{i}(x,p)}{\de x}=\frac{\de}{\de x}\left[D_{i}(x,p)\frac{\de f_{i}(x,p)}{\de x}\right]+\frac{p}{3}\frac{{\rm d} \tilde{u}(x)}{{\rm d} x}\frac{\de f_{i}(x,p)}{\de p}+Q_{i}(x,p)\,.
\label{eq:trans}
\end{equation}
Here $D_{i}(x,p)$ is the parallel diffusion coefficient, which may depend both on space and momentum, $\tilde{u}(x)=u(x)+v_W$ is the total velocity of the scattering centers in the shock frame, given by the sum of the fluid velocity $u(x)$ and wave velocity $v_{W}$, and $Q_{i}(x,p)$ is the injection term. 
The shock is at $x=0$ and subscripts 0, 1 and 2 label quantities taken, respectively, at the upstream free-escape boundary $x=x_{0}$, immediately upstream and downstream of the subshock.    

An especially important issue, when taking into account nuclei, is that of particle injection, as stated in the introduction and as should become clear below. 
Unfortunately, the microphysics of this process is not yet fully understood even for the case of protons alone, and much more so for nuclei. In the following we simply assume that protons are injected from downstream via thermal leakage as described in \citep{bgv05}, while the injection of HN is tuned in such a way as to reproduce the relative abundances observed in the CRs.
We do not account for the details of the HN injection, likely related to the complex physics of the dust sputtering process. Indeed, the required preferential injection of HN is likely related to the fact that partially ionized heavy particles (i.e., thermalized particles with large mass/charge ratios) have large Larmor radii and are hence preferentially injected in a thermal leakage scenario (see \cite{eje81}), and/or in the fact that refractory nuclei can be efficiently injected via dust grain sputtering \cite{dust2} (we will discuss these points in \S\ref{sec:abunda}).

More precisely, we assume that all protons with momentum $p>p_{inj,\prot}$ have a large enough Larmor radius to cross the shock (subshock) from downstream and start being accelerated. Since the shock thickness is expected to be of the order of the Larmor radius of particles with thermal momentum $p_{th}=\sqrt{2m_{\prot}k_{B}T_{\prot,2}}$ (where $T_{\prot,2}$ is the downstream proton temperature and $k_{B}$ is the Boltzmann constant), we take $p_{inj,\prot}=\xi_{\prot}\ p_{th,\prot}$, with $\xi \sim 3-4$. Furthermore, at a given momentum $p$ the Larmor radius of a HN with charge $Z_{i}e$ is a factor $Z_{i}$ smaller than that of a proton. Hence, in this scheme, it is very natural to assume that $p_{inj,i}=Z_{i}\ p_{inj,\prot}$.

Clearly, if nuclei are not completely ionized or if they are injected via dust sputtering this simple recipe may fail to describe the low-energy tail of the CR distribution. 
Nevertheless, since the spectra of the relativistic particles observed at Earth are consistent with power-laws extending from a few GeV/nucleon up to the knee, it seems reasonable to assume that injection always occurs at energies between thermal and mildly relativistic. Therefore, the spectra we work out are expected to be accurate throughout the entire energy region relevant for the SNR emission, and for the measured Galactic CRs as well.
 
The injection term in Eq.~\ref{eq:trans} can be written as 
\begin{equation}\label{eq:Q}
Q_i(x,p)=Q_{i,1}(p)\delta(x)=\frac{\eta_i n_{0}u_{0}}{4\pi p_{inj,i}^{2}}\delta(p-p_{inj,i})\delta(x)\,,
\end{equation}
where $\eta_i$ has the usual meaning, {\it i.e.} the fraction of particles of each species crossing the shock that is injected in the acceleration process. We will discuss in \S\ref{sec:abunda} how to determine reasonable values of $\eta_{i}$ as inferred from what is measured at the Earth. 

We solve Eq.~\ref{eq:trans} along with the spatial boundary condition $f_{i}(x_{0},p)=0$, which mimics the presence of a free escape boundary upstream placed at $x=x_{0}$.
For each chemical element, the distribution function $f_{i}(x,p)$ and the escape flux $\phi_{esc,i}(p)$ can be written as \citep[see][]{bound}: 
\begin{equation}\label{eq:app}
f_{i}(x,p)=f_{sh,i}(p)\exp\left[-\int_{x}^{0}\di x'\frac{\tilde{u}(x')}{D_{i}(x',p)}\right]	\left[ 1-\frac{W_{i}(x,p)}{W_{i,0}(p)}\right];
\end{equation}
\begin{equation}
\phi_{esc,i}(p)=-\left[D_{i}(x,p)\frac{\partial f_{i}}{\partial x}\right]_{x_0}=- \frac{\tilde{u}_{0}f_{sh,i}(p)}{W_{i,0}(p)}\,,
\end{equation}
where we have introduced the functions $f_{sh,i}(p)=f_i(0,p)$, and 
\begin{equation}\label{eq:W}
	W_{i}(x,p)=\tilde{u}_{0}\int_{x}^{0} \frac{\di x'} {D_{i}(x',p)}\exp\left[\int_{x'}^{0}\di x''\frac{\tilde{u}(x'')}{D_{i}(x'',p)}\right]\ .
\end{equation}

The distribution function at the shock reads:
\begin{equation}\label{eq:solshock}
f_{sh,i}(p)=\frac{\eta_{i} n_{0}}{4\pi p_{inj,i}^{3}}\frac{3\Rt}{\Rtt U_{p,i}(p)-1}
	\exp\left\{-\int_{p_{inj,i}}^{p}\frac{\di p'}{p'}\frac{3\Rtt}{W_{0,i}(p')}
	\frac{W_{0,i}(p')U_{p,i}(p')-1}{\Rtt U_{p,i}(p')-1}\right\},
\end{equation}
where we used
\begin{equation}\label{eq:Up}
U_{p,i}(p)=\tilde{U}_{1}-\int_{x_{0}}^{0}\di x \frac{\di \tilde{U}(x)}{\di x}\frac{f_{i}(x,p)}{f_{sh,i}(p)}
\end{equation}
and introduced the subshock $\Rs=u_{1}/u_{2}$ and the total $\Rt=u_{0}/u_{2}$ compression ratios for the fluid.
In this approach in which the waves move with velocity $v_{A}$ with respect to the fluid, the cosmic rays which scatter against them feel an effective velocity given by $u+v_{A}$ and hence a subshock and a total compression ratios given by
\begin{equation}\label{eq:rsrtt}
\tilde{R}_{sub}=\frac{u_{1}+v_{A,1}}{u_{2}+v_{A,2}};\qquad \tilde{R}_{tot}=\frac{u_{0}+v_{A,0}}{u_{2}+v_{A,2}}.\end{equation}
As we discuss below, the particles may feel compression ratios larger or smaller than the fluid, depending on the relative sign of $u$ and $v_{A}$. 
  
The convection-diffusion equation is coupled with the standard conservation equations for mass, momentum and energy flux, where the cosmic ray terms are intended to be summed over all the elements.
In particular, the conservation of momentum, after dividing by the bulk pressure $\rho_{0}u_{0}^{2}$, reads:
\begin{equation}\label{eq:mom}
U(x)+P_{g}(x)+P_{c}(x)+P_{w}(x)=1+\frac{1}{\gamma M_{0}^{2}}~,
\end{equation}
where the gas pressure, if the heating in the precursor is purely adiabatic, can be written as
\begin{equation}
P_{g}(x)=\frac{U(x)^{-\gamma}}{\gamma M_{0}^{2}}\,.
\end{equation}
If turbulent heating is also at work converting a fraction $\zeta$ of the magnetic pressure into thermal energy, the pressure of the gas would instead read \citep[see \S 4.2 of][]{lungo}:
\begin{equation}\label{eq:Pgas}
	P_g(x) \simeq \frac{U(x)^{-\gamma}}{\gamma M_{0}^{2}}\left[1+\zeta H(x)\right]\,;\qquad
	H(x)=\gamma(\gamma-1)\frac{M_0^2}{M_{A,0}}
	\left[\frac{1-U^{\gamma+1/2}(x)}{\gamma+1/2}\right].
\end{equation}

The pressure in cosmic rays is, as usual,
\begin{equation}\label{eq:Pc}
	P_{c}(x)=\sum_{i}P_{i}(x)=\frac{4\pi}{3}~\sum_{i}\int_{p_{inj,i}}^{+\infty}{\rm d}p ~p^3 ~v(p) ~f_{i}(x,p).
\end{equation}
Finally the pressure in the form of magnetic turbulence, which has been shown to play a key role in the shock dynamics \citep{jumpl}, can be approximated, assuming that only standard Alfv\'en waves are generated via resonant streaming instability and for $M_{0}, M_{A}\gg 1$, as \citep{lungo}
\begin{equation}\label{eq:Pw}
P_{w}(x)=\left( 1-\zeta\right) U(x)^{-3/2}\left[\frac{1-U(x)^{2}}{4M_{A,0}}\right]\,.
\end{equation}
A solution of the problem of NLDSA can be obtained through the same recursive method described in \cite{bound}. 
We summarize it here.
\begin{itemize}
\item Start from a guess for $U(x)$ and for the distribution function $f_{i}(x)$ ({\it e.g.} the test-particle solution) and fix a value of $U_{1}^{*}=\Rs/\Rt$. This corresponds to a value of $P_{c,1}^{*}$, through Eq.~\ref{eq:mom}, since $P_{g}$ and $P_{w}$ depend only on $U$.
\item Compute the new distribution functions $f_{i}(x,p)$ by using $U(x)$ and $f_{i}(x,p)$ from the previous step in Eqs.\ref{eq:W}-\ref{eq:Up} 
\item Compute the CR pressure through Eq.~\ref{eq:Pc}. In general, the value of $P_{c,1}$ that is obtained will be different from $P_{c,1}^{*}$. Then $P_{c}(x)$ and $f_{i}(x,p)$ are {\it renormalized} by multiplying them by a factor $\lambda=P_{c,1}^{*}/P_{c,1}$.
\item Compute and update the fluid velocity profile $U(x)$ using the new $P_c(x)$ in Eq.~\ref{eq:mom}.
\item Iterate the steps above until $\lambda$ does not change between a step and the following.  
\end{itemize}
Starting from an arbitrary $U_{1}^{*}$, in general one has $\lambda\neq 1$ after convergence is reached. The process must then be restarted with a new choice of $U_{1}^{*}$ until convergence to $\lambda=1$ is achieved (within a prescribed level of accuracy). 
The fluid profile and the distribution functions so obtained are, at the same time, solutions of both the convection-diffusion and the conservation equations, {\it  i.e.} solutions of the full problem. 

It is worth recalling that from the computational point of view this approach to non-linear shock acceleration is the most efficient presented so far in the literature: for a given set of parameters it takes less than one minute (on a laptop) to calculate the distribution function $f_{i}(x,p)$ of all accelerated particles and the space dependent structure of all thermodynamical quantities in the shock region.

\section{Abundances of heavy elements}\label{sec:abunda}

The most challenging aspect of a non-linear calculation of diffusive shock acceleration in the presence of accelerated nuclei is the tuning of the relative abundances at the source in a way that may fit the abundances observed at the Earth after propagation. The difficulty is manyfold: first, the non-linearity of the problem makes it very difficult to establish once and for all what the required relative abundances are; second, the spectrum observed at Earth from an individual SNR is the superposition of the instantaneous spectra of particles escaping the remnant at different times, and the relative abundances are, most likely, time dependent. One final complication to keep in mind in establishing a connection between source and local abundances of ions comes from the spallation processes suffered by nuclei during propagation in the Galaxy.
This simply means that the relative abundances of chemicals in a given remnant at a given time cannot be univocally inferred from the CR abundances that we measure at Earth.

It is worth recalling that the situation is rather different for the EBS, where real-time measurements of the spectra of different elements are available, thus allowing a more detailed interpretation of the ongoing physical processes. In this case, in fact, the injection of both protons and nuclei via thermal leakage provided a successful explanation of the EBS properties as due to efficient NLDSA \cite{emp90}. 

On the other hand, the same thermal leakage scheme seems unsatisfactory when applied to the case of SNRs as the sources of Galactic CRs.
In fact, a thorough study of the elemental and isotopic ratios in CRs compared with solar ones shows that refractory elements are preferentially accelerated with respect to volatile ones. We refer to the original works \cite{dust1, dust3} (and to a recent review \cite{wiede}) for a comprehensive discussion on these topics.

Here we only want to stress that this fact has been explained by invoking the fundamental role of the interstellar dust in the injection of nuclei, as put forward in \cite{dust2}. In few words, when a grain of refractory material is caught by a SNR shock, it is efficiently accelerated by crossing the shock many times thanks to its large Larmor radius. When its velocity reaches about $0.01c$ it begins to sputter and to release atoms. These have the same velocity of the parent grain: they are hence well suprathermal and therefore can be easily injected and take part in the acceleration process. 
Such an effect can also account for the fact that among refractory nuclei there is no preferential acceleration of heavy elements with respect to lighter ones, which is instead the case for volatile elements. This latter evidence is usually explained as a consequence of the fact that heavy elements may be ionized only partially and hence preferentially injected because of their relatively large rigidity (see \cite{david79,eje81}).

In this section we illustrate for simplicity how to estimate the relative normalization between source spectra of different ions in a test-particle approach. 
Such an estimate is expected to hold for energies above a few GeV/nucleon, where the spectra detected at Earth are very close to power-laws. The situation may be more complicated in the region between slightly suprathermal and weakly relativistic particles because of the effects of partial ionization and dust sputtering discussed above.

The resulting abundances, not too dissimilar from the exact ones derived in \S \ref{sec:allpart}, are used to make several important points, mainly concerning the instantaneous spectra of relativistic particles (for instance to address the dynamical role of nuclei at the shock, and their effect on the generation of secondary radiation). This discussion can be carried out regardless of a detailed knowledge of the non-relativistic region of the spectra.

The starting point of our estimate is represented by the observed relative abundances in CRs at Earth \citep[see e.g.][for a recent review of the observational results]{hoer08}. In a test-particle approach the spectrum of particles accelerated at the shock can be written as a power law with slope $\beta$:
\begin{equation}
f_{i}(p)=\frac{C\eta_{i}}{4\pi p_{inj,i}^{3}}\left(\frac{p}{p_{inj,i}}\right)^{-\beta},
\label{eq:fi}
\end{equation}
with $C$ the same constant for all $i$. Let us also assume $p_{inj,i}=\alpha_{i}\ p_{inj,\prot}$. In the following we shall assume that the injection is charge dependent, with $\alpha_{i}=Z_{i}$, being $Z_{i}e$ the electric charge of ions of type $i$. The number of particles with momentum $>p$ can be estimated as $F_{i}(>p)\approx 4\pi p^{3} f_{i}(p)=C \eta_{i} (p/p_{inj,i})^{3-\beta}$. The spectrum at Earth is affected by propagation as
\begin{equation}
n_{i}(p)\approx F_{i}(p) \tau_{i}(p) = C \eta_{i} \left(\frac{p}{p_{inj,i}}\right)^{3-\beta}  Z_{i}^{\delta} \tau_{\rm H}(p),
\label{eq:conf}
\end{equation}
where we used the fact that the confinement time $\tau_{i}$ of nuclei of charge $Z_{i}$ is $Z_{i}^{\delta}$ times larger than that of protons at given momentum. This reflects the rigidity dependent nature of particle propagation in the Galaxy, namely the fact that the diffusion coefficient is $D(p)\propto R^{\delta}=(p/Z_{i})^{\delta}$ ($R$ is the particle rigidity). The simple scaling in Eq. \ref{eq:conf} holds as long as spallation in the ISM can be neglected, therefore it should only be used at sufficiently high energies (we use $p_{*}=10^{5}$ GeV/c as normalization point) so that the escape time from the Galaxy is much shorter than the spallation time scale for all the relevant species. 

The ratio of abundances between ions and protons at the same momentum $p_{*}$ as measured at the Earth is therefore:
\be
K_{i{\rm H}}=\frac{n_{i}}{n_{\rm H}} = \frac{\eta_{i}}{\eta_{\rm H}}  Z_{i}^{\delta} \left( \frac{p_{inj,\rm H}}{p_{inj,i}} \right)^{3-\beta} = 
\frac{\eta_{i}}{\eta_{\rm H}}  Z_{i}^{\delta+\beta-3}.
\ee
In this expression, the ratio $K_{i{\rm H}}$ comes from the measured spectra and, at least in the context of this simple (leaky-box-like) approach, $\beta+\delta$ is fixed by observations to be $\sim 4.7$. The ratio of the $\eta$'s to be used in the calculations can therefore be easily estimated from measurements. The values resulting from this procedure are illustrated in Tab.~\ref{tab:kappa}. In the same table we also show the flux of different chemicals at 1 TeV and the spectral slope at Earth as inferred using the so-called Poly-Gonato parametrization by H\"{o}randel \cite{hoer03}. These slopes reflect the presence of spallation and most likely the superposition of different types of sources (see discussion in \S \ref{sec:discuss}). 

\begin{table}
\centering
 \begin{tabular}{@{}lcccc}
  \hline
  Element  & Flux(1 TeV) & $\gamma_{i}$ & $K_{i{\rm H}}$ & $\frac{\eta_{i}}{\eta_{\rm H}}$\\
  \hline
  H		& $(8.73\pm 0.07)\times 10^{-2}$		&	$2.71\pm 0.02$	& 1 & 1\\
  He		& $(5.71\pm 0.09)\times 10^{-2}$		&	$2.64\pm	0.02$	&	0.90 & 0.28\\
  C-N-O$^{(*)}$   & $\sim 5.4\times 10^{-2~(*)}$		&	$\sim 2.64^{(*)}$	&	0.85 & 0.03\\
  Mg-Al-Si & $\sim 1.7\times 10^{-2}$		&	$\sim 2.66$	&	0.25 	& $3.2\times10^{-3}$\\
  Fe		&	$(2.04\pm 0.26)\times 10^{-2}$		&	$2.59\pm 0.06$		&	0.41 &	$1.6\times 10^{-3}$\\
  \hline
 \end{tabular}
\caption{Flux at 1 TeV (in units of [m$^{2}$ sr s TeV]$^{-1}$), spectral slope ($\gamma_{i}$) and ratio relative to H at $10^{5}$ GeV  ($K_{i \rm{H}}$) measured at Earth for the most abundant elements in galactic CRs. 
CR data are from \cite{hoer03}, table 7, except the ones denoted with $(*)$ which are from \cite{cream10}. C, N and O are taken as one effective element with $Z_{\rm CNO}=7$ and $A_{\rm CNO}=14$. The same is done for Mg, Al and Si, with $Z_{\rm MgAlSi}=13$ and $A_{\rm MgAlSi}=27$. \label{tab:kappa}}
\end{table}
  
As stressed above, the procedure just outlined is based upon a test-particle picture and should only be used as a qualitative estimate of the efficiency of acceleration of the different chemicals. Many factors can affect the result of this simple estimate: (1) the non-linear diffusive acceleration at shocks does not lead to power law spectra; (2) the values of $\eta_{i}$ can be (and in general are expected to be) time dependent, which is especially relevant since the spectrum of CRs observed at Earth results from the superposition of instantaneous spectra of particles escaping the SNR at different times; (3) there are different types of SNRs exploding in different environments, which leads to different ratios at the sources; (4) spallation changes the spectra in a complex way, and the difference in the spectral slopes of different chemicals as reported in the table above can partially be due to these reactions, although it is likely that the differences also reflect the contributions from SNRs in different ambient media; (5) the diffusion coefficient is not strongly constrained by the secondary to primary ratios and the slope $\delta$ could well be in the range $0.3-0.7$, though the upper end of this range leads to problems with anisotropy (e.g. \cite{hillas05}); (6) although the spectrum observed at Earth is very close to a power law, different SNRs or the same remnant at different stages of its evolution may have both different spectra and different values of the relative abundances.

In \S\ref{sec:spectra} and \S\ref{sec:gamma} we adopt the simple procedure outlined above to estimate the abundances of ions in a source at a given time, while in \S\ref{sec:allpart}, where we calculate the spectrum of the different chemicals at the Earth and the all-particle spectrum, we follow a somewhat more refined procedure. It is however wise to keep in mind that, for all the reasons mentioned above, it is extremely difficult, if at all possible, to infer in a realistic way the abundances at the sources that may fit the whole set of data available at Earth. 

\section{Particle spectra and shock hydrodynamics}
\label{sec:spectra}

We consider a shock moving with velocity $u_{0}=4000$ km/s in a homogeneous medium with particle density $0.01$ cm$^{-3}$, temperature $T_{0}=10^{6}$K (sonic Mach number $\sim 34$) and magnetic field $B_{0}=5\mu$G aligned with the shock normal. This choice of parameters corresponds to a 2000 year old SNR with radius $R_{sh}\sim 14.4$ pc, i.e. a SNR at the beginning of its Sedov-Taylor stage for a SN explosion of $10^{51}$ erg and an ejecta mass of 1.4 solar masses. It is worth recalling that the highest cosmic ray energy is thought to be achieved at this evolutionary stage \citep{bac07}.

The free-escape boundary is placed at $x_{0}=0.2R_{sh}$ upstream of the shock and the diffusion coefficient is taken as Bohm-like, 
\begin{equation}
D_{i}(x,p)=\frac{1}{3}v(p)\frac{pc}{Z_{i}B(x)},
\end{equation}
in the amplified magnetic field at the shock position, namely $B(x)=B_{1}=\sqrt{8\pi\rho_{0}u_{0}^{2}P_{w,1}}$ upstream and $B(x)=B_{2}=\Rs B_{1}$ downstream.

\begin{figure}
\centering
\includegraphics[scale=0.32]{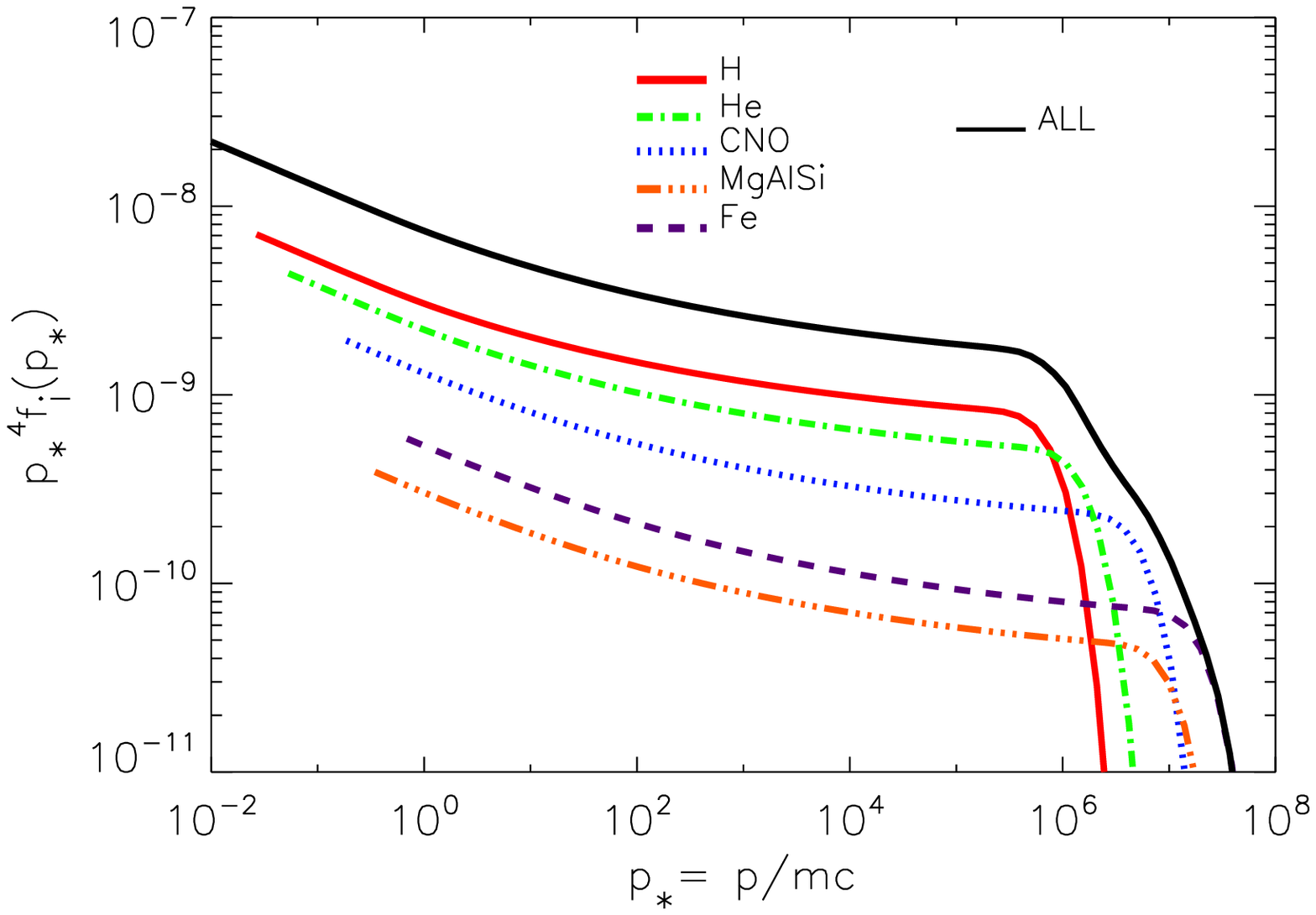}
\includegraphics[scale=0.32]{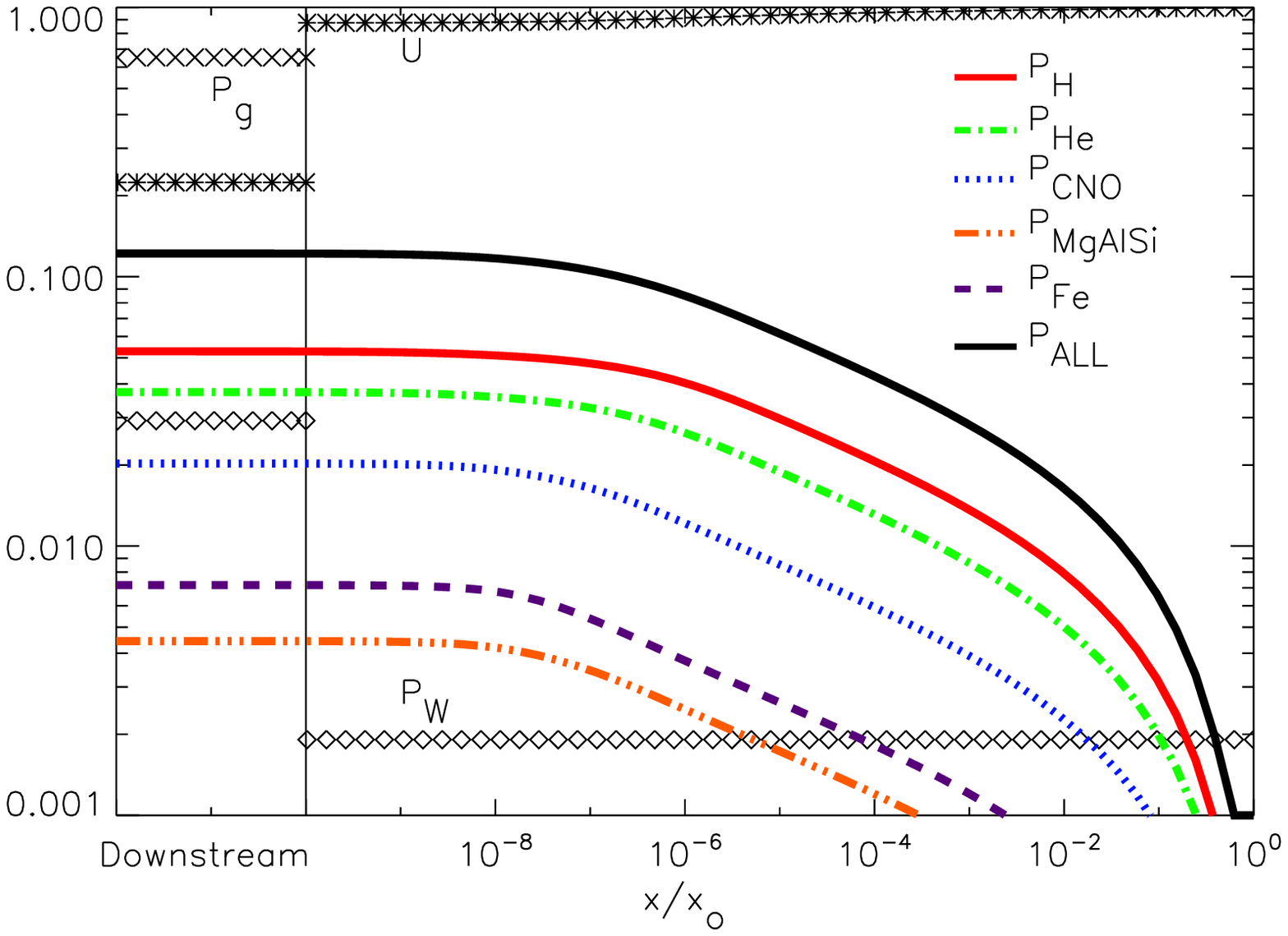}
\caption{\emph{Left panel}: all-particle spectrum (thick line) and spectra of individual elements. \emph{Right panel}: spatial dependence of the hydrodynamical quantities (the gas pressure upstream is very low and lies outside the plot boundaries).}
\label{fig:spectra}
\end{figure}

In Fig.~\ref{fig:spectra} we show the spectra of accelerated particles and the the quantities related to shock hydrodynamics, obtained through the iterative method described in \S\ref{sec:model}, in a case of efficient particle acceleration (we used $\xi_{\prot}=3.8$, corresponding to $\eta_{\prot}=5.7\times 10^{-5}$ in Eq.~\ref{eq:Q}). Notice that the gas pressure upstream is very low and lies outside the plot boundaries.

The most noticeable feature is the fact that, for the standard abundances deduced in \S\ref{sec:abunda}, the dynamical role of nuclei overall is twice as important as that of protons: at the shock position the pressure of accelerated protons is $P_{\prot}\simeq 0.05$, in units of the ram pressure far upstream, while the pressure in the form of relativistic HN is about 0.1 (right panel of Fig.~\ref{fig:spectra}). The latter is mainly associated to He nuclei ($P_{\he}\simeq 0.045$) but also CNO and Fe nuclei carry non-negligible fractions of the momentum in the form of accelerated particles ($P_{\cno}\simeq 0.017$ and $P_{\fe}\simeq 0.003$).
The relative importance of HN with respect to protons should not be affected by the exact shape of the spectra between $p_{inj,i}/Z_{i}$ and a few GeV in rigidity (which is related to the injection details discussed above), provided that the spectra are not much steeper than $p^{-4}$ and/or that the maximum momentum is larger than few GeV/c. These peculiar conditions might only be met rather late in the remnant evolution, and definitely not as early as at the beginning of the Sedov stage, i.e., when the acceleration efficiency is at its maximum.

For comparison, in the right panel of Fig.~\ref{fig:spectra} we also show the pressure associated to the thermal plasma and to the magnetic turbulence (taken as homogeneous upstream) and the normalized fluid velocity as well. 
The latter shows the spatial profile typical of cosmic ray modified shocks, with a precursor, a weak subshock with compression ratio $\Rs\simeq 3.9$ and a total compression ratio larger than 4, namely $\Rt\simeq 4.5$.

The resulting spectra of the accelerated particles reflect the modified dynamics, being steeper than the test-particle prediction, $\sim p^{-4}$, at low energies, and flatter than $p^{-4}$ at the highest energies. The finite shock size (spatial boundary), which allows for particle escape from upstream, induces a rigidity-dependent cut-off in the spectra of the various species, leading to maximum momenta scaling with atomic charge (left panel of Fig.~\ref{fig:spectra}).
More precisely, since the diffusion is rigidity-dependent, and since in diffusive shock acceleration the spectral slope at a given momentum, $q_{i}(p)=-\frac{{\rm d}\log f_{i}(p)}{{\rm d}\log p}$, depends only on the compression ratio actually experienced by the particles with that momentum, we have the simple scaling: $q_{i}(p/Z_{i})=q_{\prot}(p)$.  
In principle, when the relative normalization and the ionization state are fixed, and when the CR pressure is dominated by relativistic particles, one could use this simple scaling to calculate the distribution functions of all the elements at the shock, their spatial dependence via Eq.~\ref{eq:app} and finally the total pressure in cosmic rays $P_{c}(x)$ and the whole shock structure.
However, within the fast semi-analytic formalism outlined in \S\ref{sec:model}, the computational effort is far from being a severe issue demanding numerical optimization.

For comparison, we solved the set of equations describing a shock with exactly the same characteristics, except for the fact that no HN were considered. We found $P_{\prot}\simeq 0.07$ at the shock. Comparing this result with those in the right panel of Fig.~\ref{fig:spectra}, the conclusion is that when acceleration of HN is taken into account, the pressure in accelerated protons is decreased, but the total pressure in accelerated particles is larger, and the shock correspondingly more modified. Indeed, in the absence of HN, the compression ratios turn out to be $\Rs\simeq 3.95$ and $\Rt\simeq 4.2$.  At the same time, also the magnetic field amplification via resonant streaming instability of the accelerated particles gives different results when the acceleration of HN is included. We find that the proton-only case returns a downstream magnetic field of $\sim 33\mu$G, while with HN $B_{2}\simeq 47\mu$G. This result is easily understood given the dependence of the amplified magnetic field on the pressure of accelerated particles. 
All these effects contribute to make the case for the necessity of taking HN into account properly: the acceleration of HN cannot be treated  in a test-particle approximation, nor can it be linearly added to a proton only case. In order to correctly describe the shock, all species need to be taken into account in a fully non-linear calculation.

A crucial point to stress is that the mild shock modification and the correspondingly steeper spectra found here and illustrated in Fig. \ref{fig:spectra} are the consequence of the assumption of scattering centers moving with a wave velocity that equals the Alfv\'en velocity calculated in the amplified magnetic field. As discussed by Caprioli et al. \cite{crspec}, without this assumption the spectra produced in SNRs are very flat and can hardly be related to the CR spectrum observed at Earth, unless, as done by Berezhko and V\"{o}lk \cite{berevolk}, one assumes a very strong dependence of the diffusion coefficient in the Galaxy on energy, $D(E)\propto E^{0.75}$, which however leads to too large anisotropy at the knee compared with observations. It is therefore of the highest importance to realize how the main result in this type of calculations is actually due to one of the aspects that are least known. 

The fact that taking into account the velocity of the scattering centers could lead to steeper spectra was already recognized in the early literature on the topic \citep[see e.g.][]{bell78}, but the issue has become much more discussed in recent times, after discovery of highly amplified magnetic fields in the shock region. There are two levels of the problem: (1) we are unable to determine the velocity of the waves responsible for the scattering from first principles, especially in the case of relevance for us in which magnetic field is strongly amplified by CR induced instabilities. If one assumes (unrealistically) that even in these extreme circumstances the waves remain Alfv\'en waves, namely with magnetic field perpendicular to the (much smaller) background magnetic field, then the Alfv\'en speed is well defined, but its value is so small to induce negligible effects on the spectrum of accelerated particles. On the other hand, one might expect that an effective value of the wave speed is close to the  Alfv\'en speed calculated using the strength of the amplified field as reference value. In this case the Alfv\'en speed is much larger and its effect on the transport equation is not negligible. (2) Even if we knew how to calculate the wave velocity (for instance by assuming that it corresponds to the Alfv\'en speed in the amplified field), the helicity of the waves is unknown, being related to the type of wave, the mechanism that generated them and the transport of the waves through the shock (reflection and transmission). In other words, it is not easy to calculate the wave velocity with respect to the shock surface.

Since the slope of the CR spectrum depends on the compression ratio of the scattering centers, as felt by the particles, it is clear from Eq.~\ref{eq:rsrtt} that, when $v_{W}$ is not negligible compared with $u$, the actual spectrum may be flatter or steeper than the standard prediction depending upon the relative sign of $u$ and $v_{W}$, both in the upstream and in the downstream regions. If $v_{W}$ is calculated in the amplified magnetic field, it typically turns out to be a not-neglibile fraction of $u$, and thus the helicity of the waves strongly affects the shape of the CR spectrum.

While upstream the turbulence is very likely generated by the CR gradient, and thus $v_{W}$ is expected to have sign opposite to $u$, downstream the waves may retain the same helicity (as one would expect based on the transmission and reflection coefficients at the subshock discontinuity appropriate for Alfv\'en waves), or rather may be fully isotropized ($v_{W}=0$). Another possibility, recently explored by Ptuskin et al. \cite{pzs10}, is that downstream waves may also be generated via streaming instability by the advected CRs, whose gradient is induced by the postshock evolution and in particular by adiabatic losses in the expanding shell: this scenario leads these authors to assume that in the downstream $v_{W}$ has the same sign as $u$. This mechanism may be at work in SNRs, but one would expect that it should act in the downstream plasma on hydrodynamical spatial scales. It is therefore hard to assess the relevance of this process as compared with that of the turbulence produced upstream and then advected through the subshock in the downstream region.
In this work we conservatively assume that downstream $v_{W,2}=0$, which leads to a less dramatic dependence of the result on the Alfv\'enic Mach number with respect to the assumptions of \cite{pzs10}, even if in both cases $M_{A}$ is calculated in the amplified magnetic field.

Moreover, in our model the strength of the amplified magnetic field is not fixed a priori but is rather a non-linear output of the full calculation, depending on the efficiency of the CR acceleration through Eq.~\ref{eq:Pw}. In this sense our approach includes a self-regulating effect which returns almost the same result in terms of particle spectra and shock modification for a wide range of values of the Alfv\'enic Mach number at upstream infinity, $M_{A,0}$, and of the particle injection efficiency $\xi_{i}$.

\section{$\gamma-$Rays from accelerated nuclei}
\label{sec:gamma}
In the typical environment of a SNR, in addition to playing an important role on the shock dynamics, accelerated nuclei may also give a contribution in terms of $\gamma-$ray emission via $\pi^{0}$ production in nuclear interactions with thermal protons.
In order to calculate this emission, we model each accelerated nucleus with atomic number $A_{i}$ and energy $E$ as an ensamble of $A_{i}$ protons, each with energy $E/A_{i}$, and adopt the parametrization of the nuclear proton-proton cross section worked out by Kamae et al. \cite{kamae+06}. 

A simple estimate of the HN contribution to pion production can be obtained as follows.
Let us assume that a nucleus with energy $E_{i}$ produces monochromatic photons with energy $E_{\gamma}$ such that $E_{i}=\chi A_{i} E_{\gamma}$, with $\chi\approx 10$. 
The photon spectrum produced by nuclei of species $i$ would thus be: 
\begin{equation}
N_{i}^{\gamma}(E_{\gamma})\propto A_{i} N_{i}(E_{i})\frac{\di E_{i}}{\di E_{\gamma}}= \chi A_{i}^{2} N_{i}(E_{i})\,. 
\end{equation}
For spectra $N_{i}(E_{i})\propto E_{i}^{-q}$ up to an energy $E_{max,i}$ we have:
\begin{equation}
\frac{N_{i}^{\gamma}(E_{\gamma})}{N_{\prot}^{\gamma}(E_{\gamma})}=K_{i} A_{i}^{2-q}\,,
\end{equation}
which means that, for the standard case $q=2$, all the species contribute to the $\gamma$-ray flux proportionally to their abundances at any photon energy below the minimum of $E_{max,i}/(A_{i}\chi)$. 
Since typically $E_{max,i}\propto Z_{i}$, the photon spectrum of any HN is cut-off at an energy which is a factor 2 lower than for protons.
It is also interesting to notice that, if the accelerated particle spectra are flatter than $E^{-2}$, the contribution of HN with respect to protons is boosted by a factor $A_{i}^{2-q}$. This turns out to be as large as a factor $\sim 5$ in the case of Fe nuclei accelerated with a spectrum $\sim E^{-1.5}$. There may be stages in the evolution of a SNR when the shock is strongly modified and such hard spectra may appear.

\begin{figure}
\centering
\includegraphics[scale=0.50]{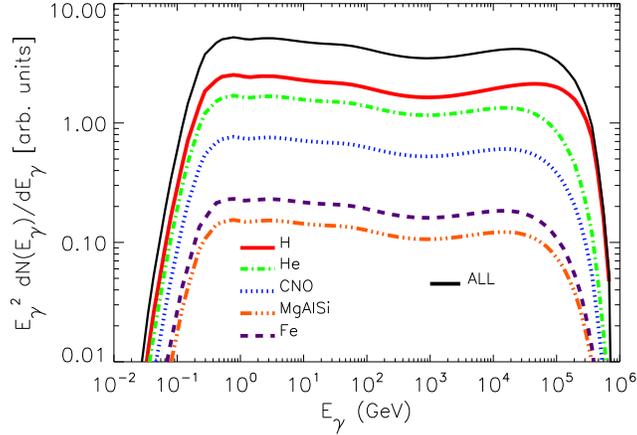}
\caption{Spectra of $\gamma$-rays due to the decay of $\pi^{0}$s produced in nuclear interactions (thick line). The partial contributions by each element are also shown.}
\label{fig:gamma}
\end{figure}

The spectrum of $\gamma-$rays, calculated using the particle spectra shown in Fig.~\ref{fig:spectra}, is illustrated in Fig.~\ref{fig:gamma}. The HN contribution to the $\gamma-$ray flux turns out to be dominant over that of protons, with a total predicted flux which is about a factor 2.5 larger than the standard prediction for the case when accelerated particles are only protons. Moreover, since the contribution of all species of HN is truncated a factor 2 below that obtained with protons alone, the shape of the high-energy end of the photon spectrum is somewhat different when HN are taken into account. Although this may in principle represent a spectral feature flagging for the presence of accelerated nuclei, we find it unlikely to be detectable, given the many intrinsic systematics of experimental and theoretical nature and given the topology of the emitting regions which in general are complex, so that different regions of the remnant (where the highest energy could be slightly different) contribute to the flux along the same line of sight.

For the modeling of the gamma ray emission from SNRs a crucial parameter is the gas density in the shock region. Moreover,  in the context of non-linear theory, the temperature of the downstream plasma is an output of the problem and the thermal emission from downstream (continuum due to bremsstrahlung and lines from non-equilibrium ionization) can be calculated. The latter, scaling with the square of the plasma density, is more sensitive than the gamma ray emissivity to the value of the gas density. In some cases the gas density is inferred from gamma ray observations, with the strong assumption that the gamma ray emission is due to production and decay of neutral pions and checked versus the signal in the form of lines of thermal origin  \citep[see e.g.][for the case of RX J1713.7-3946]{ellison+10}. The results shown in this section illustrate the fact that the inclusion of ions in the calculations of the gamma ray emission from a SNR leads to an estimate of the ambient density which may easily be a factor $\sim 2-4$ smaller than when the gamma ray emission is calculated using only protons. The thermal emission decreases correspondingly by $\sim 4-16$. Claims on the detectability or non-detectability of thermal emission should take into account the intrinsic uncertainty associated with this phenomenon.

\section{The Galactic spectrum of CRs}
\label{sec:allpart}

In this section we calculate the total spectrum of CRs accelerated by a class of benchmark SNRs during the different stages of evolution, as described by the analytical evolutionary scheme of \cite{TMK99}. Our reference SNR has the same environmental parameters as in \S\ref{sec:spectra}, and we follow its evolution from 0.1 to about 22$T_{ST}$, where $T_{ST}\simeq 2000$yr is the beginning of the Sedov-Taylor stage. We assume that the ambient ISM is homogeneous, hot and rarefied ($T_{0}=10^{6}\degK$, $n_{0}=0.01$) in order to mimic the explosion of a core-collapse SN into the hot bubble excavated by its pre-SN wind, since about 85\% of Galactic SNe are expected to be of this type \citep[type II + type Ib/c, see e.g.][]{hl05}. Such a choice should also be representative of the typical hot medium inside superbubbles, where most \citep[about 75\%, according to][]{hl05} of the core collapse SNe are found.
We do not account here either for the presence of a dense, cold wind, occasionally produced during the Red Supergiant stage of the progenitor, or for the finite size of the hot bubble. 
These ingredients add to the problem of particle acceleration numerous difficulties and require the introduction of several almost unconstrained parameters, like for instance the mass-loss rate and the total mass blown in the wind, the temperature and the density profile and, more important, the strength and the topology of the magnetic field.
By the same token, here we did not include the contribution of type Ia-like SNe, which are expected to be less frequent and to explode into the denser and colder homogeneous ISM.

It is important to stress that our decision to not include these situations in our calculations is not due to technical problems of our computation technique, but only to our belief that the additional uncertainties that would follow overcome the benefits of adding a class of sources to our calculations, thereby making the results not more, but in fact less trustworthy. 
Nevertheless, some of these pieces of information may be available when dealing with a given SNR: in this respect, our approach could be easily modified in order to include a more complex evolution of the remnant accounting also, e.g., for the presence of either a dense Supergiant wind or the final encounter between the forward shock and the ordinary ISM (see for instance \cite{pz05} for an analytical treatment of the evolution of different SNR types).

\begin{figure}
\centering
\includegraphics[scale=0.9]{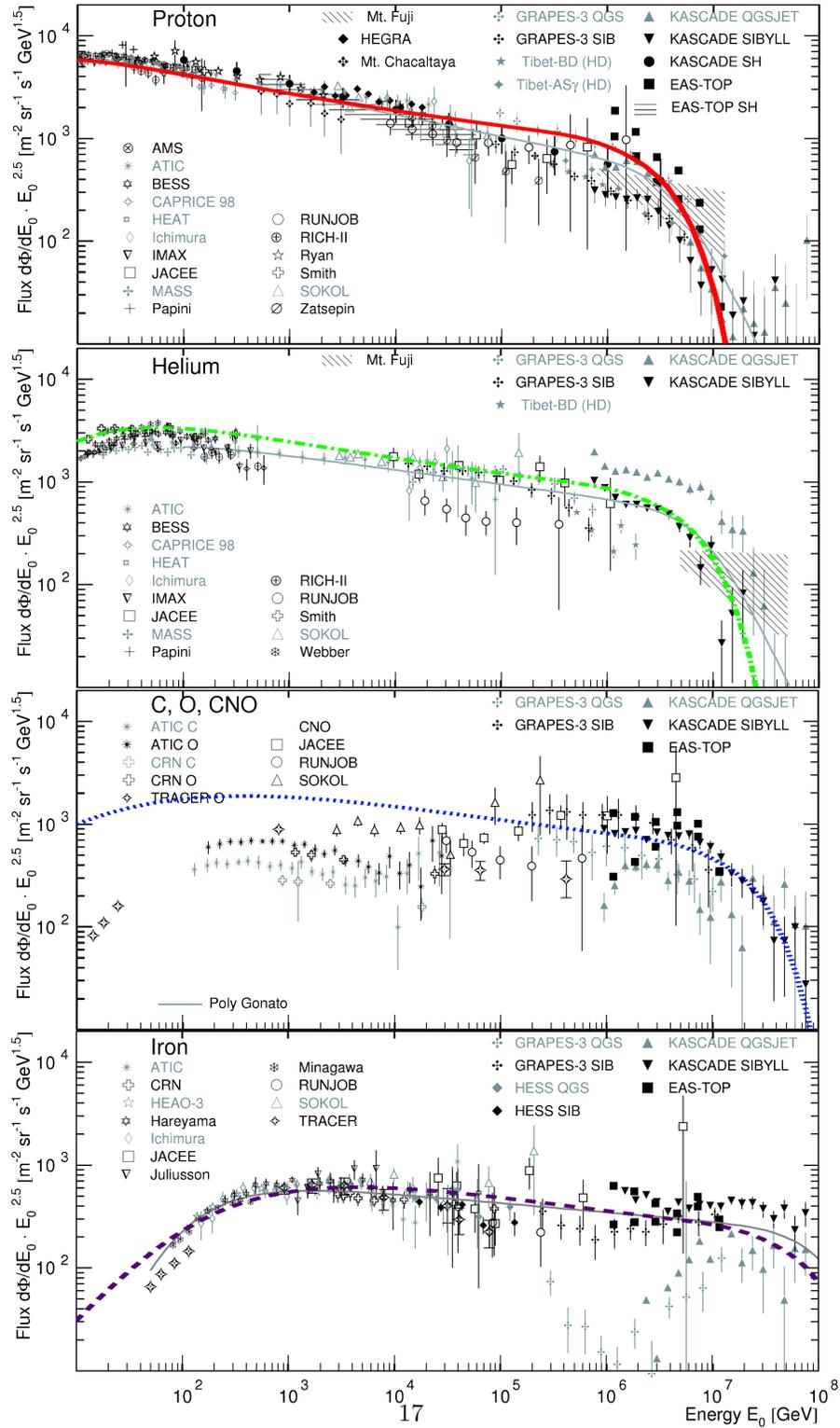}
\caption{CR flux measured at Earth for the different chemicals considered in the text \citep[the experimental data are from fig.~$9$ of][where the proper references to the sigle experiments can be found as well]{blumer+09}.
The thin solid lines in the H, HE and Fe panels correspond to the Poly-Gonato fits \citep[][,see also \ref{tab:kappa}]{hoer03}.
\label{fig:spectraALL}
}
\end{figure}

The spectra of the accelerated particles are calculated as described in \cite{crspec}, i.e. taking into account the instantaneous escape flux from the upstream boundary \citep[see][]{escape}, the advection in the downstream region, which leads to adiabatic losses as a consequence of the shell expansion, the escape of particles from downstream because of shell breaking and/or inhomogeneities in the circumstellar medium \citep[the fraction of downstream escaping particles is taken as 10\%: see also \S3.3 of][for a discussion of this point]{crspec}. The relative abundances at the sources are iteratively adjusted in order to fit the fluxes measured at Earth. 

The damping of the magnetic field is heuristically taken into account by assuming that a fraction $\zeta$ of the generated turbulence is damped into gas heating (Alfv\'en heating), with $\zeta(t)=1-\exp\left[-u_{0}(t)/u_{\rm damp}\right]$, where the velocity $u_{\rm damp}$ at which the effect becomes relevant may range between about 200 and 1000 km/s, depending on the details of the turbulence generation and damping \citep[see][for a study of this topic]{pz05}. We checked {\it a posteriori} that our findings depend only weakly on the choice of $u_{\rm damp}$.

The convolution over time of the spectra of accelerated particles injected into the Galaxy returns the source spectrum, which has been corrected by accounting for the propagation in the Milky Way. We adopt a simple leaky box model of the Galaxy, which is taken as a cylinder with radius and half-heigth equal to 10 and 3.5 kpc respectively, where 3 SN explode each century. The escape time is a function of the nucleus rigidity $R$ as inferred by standard secondary-to-primary measurements and reflects into a grammage $\lambda_{esc}(p)= 7.3 (p/Z/10 {\rm GV})^{-\delta} \beta(p)$ g cm${}^{-2}$, with $\delta=0.55$ and $\beta(p)$ the dimensionless speed of a nucleus of momentum $p$. Particle losses due to spallation against interstellar nuclei during the propagation is also taken into account and described as in \S4 of \cite{hoer+07}.

\begin{figure}
\centering
\includegraphics[scale=0.75]{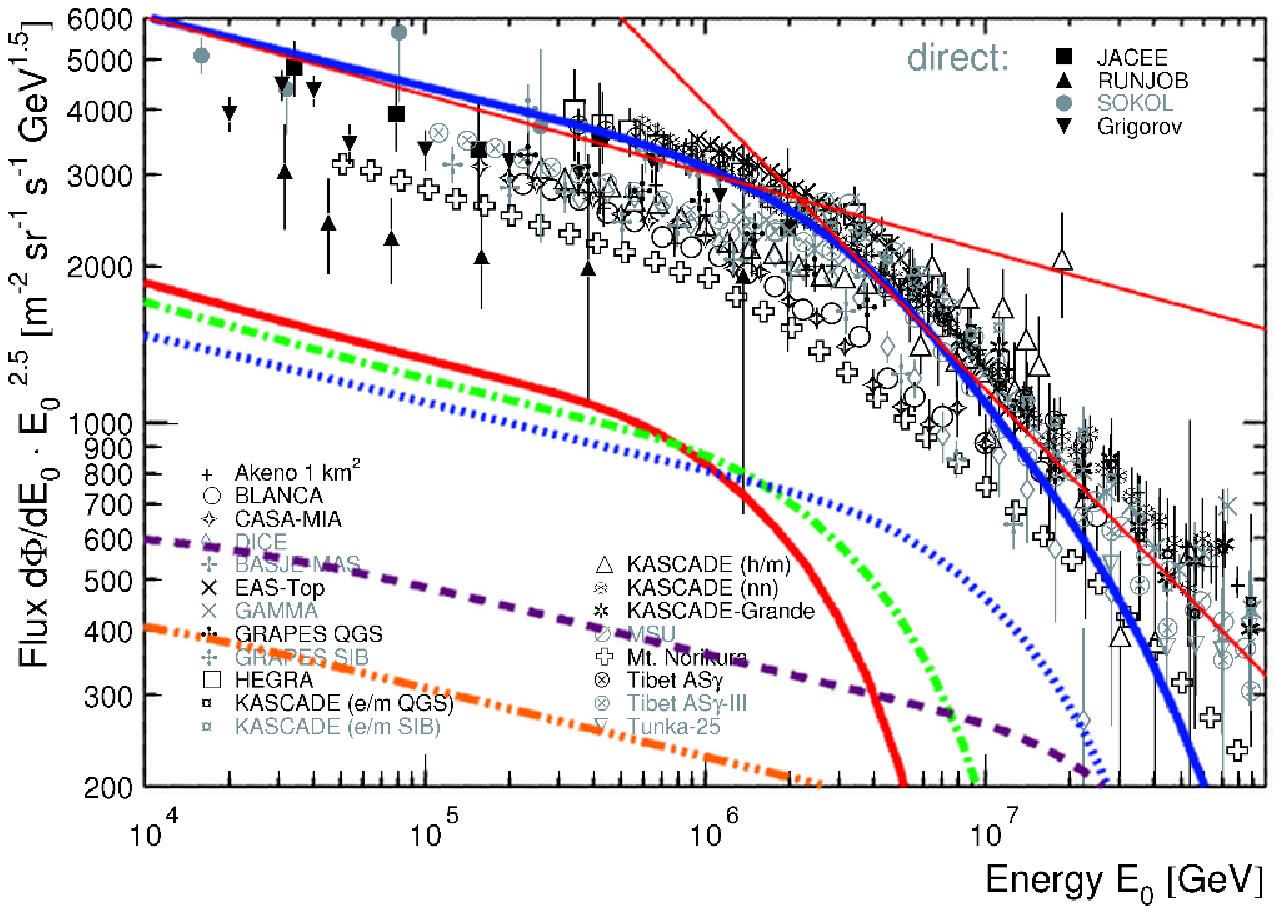}
\caption{All-particle CR flux at Earth in the knee region. The thick line represents our model output, while the thin lines guide the eye by showing two power laws, $\propto E_{0}^{-2.7}$ and $\propto E_{0}^{-3.1}$ below and above the knee, respectively \citep[data from][fig.~8]{blumer+09}. The partial contributions from H, He, CNO, MgAlSi and Fe are also shown.}
\label{fig:knee}
\end{figure}

The results for the different chemicals are shown in Fig.~\ref{fig:spectraALL}, as plotted on top of the experimental data collected in \cite{blumer+09}. The agreement is very good, and both the spectra and the absolute and relative normalization are consistent with the data from direct and indirect experiments able to resolve the chemical composition of the CR flux at Earth. Notice that some discrepancies at low energies are due to the fact that the results of our calculations have not been corrected for solar modulation. 

It is interesting to notice that the spectra of nuclei at Earth have roughly the same slope at given rigidity, except for the effect of spallation, which produces the low energy turnover. For iron nuclei the flattening due to spallation is visible up to energies of $\sim 10^{5}$ GeV. This conclusion should however be taken with caution: the recent data from the balloon-borne experiment CREAM-II \citep{cream10} suggest that spectra of nuclei are somewhat flatter than the proton spectrum. This might be a mild signature of the contribution of different classes of sources operating in environments with different chemical composition and contributing to the CR spectrum in a different way than described here.

In Fig.~\ref{fig:knee} the all-particle spectrum in the knee region is shown (thick line), again compared with data from \cite{blumer+09}. For comparison two power-laws are also drawn (thin lines), corresponding to the fiducial slopes of the flux below ($E_{0}^{-2.7}$) and above ($E_{0}^{-3.1}$) the knee.

The all-particle spectrum that we calculated is, as expected, somewhat steeper than the standard best fit to the data above the knee ($\propto E_{0}^{-3.1}$). For this effect, also found by other authors, several possible explanations have been proposed: in \S2 of the review by Hillas \cite{hillas05} the author discussed the need for an ``extended tail'' in the Galactic flux in order to explain the CR data in the region between $10^{7}$ and $10^{9}$ GeV (see fig.~2 of that paper), and suggested that such a high-energy feature might be granted by a proper account of a peculiar class of Type II SNe. A similar scenario has been put forward by Ptuskin et al. \cite{pzs10}, who successfully explained this region of the spectrum with rare (only 2\% of all Galactic SNe) but very energetic ($E_{SN}=3\times 10^{51}$ erg) type IIb SNe. As an alternative, \cite{hoer08} suggested the importance of nuclei heavier than iron, which are accelerated up to energies $\sim 100$ ($Z_{\rm U}=92$) larger than that of protons.

We wish to stress however that the spectrum of CRs in this energy region is affected by the relative fluxes of Galactic and extragalactic CRs, and this makes the claim for excesses rather weak. The only point that appears rather strong in terms of the view of the origin of CRs illustrated in the papers listed above, as well as in the present paper, is that the spectrum of Galactic CRs is unlikely to extend to very high energies, thereby making the transition from Galactic to extragalactic CRs at the ankle rather poorly motivated. The two physical scenarios which are compatible with a transition in the energy region around $10^{7}-10^{9}$ GeV are the mixed composition scenario \cite[]{allard1,allard2,allard3,allard4} and the dip scenario \cite[]{dip1,dip2,dip3,dip4}. The former model describes the extragalactic CR contribution as the superposition of different nuclei, so that the chemical composition in the transition region is mixed. The latter model is stunning for its simplicity: the extragalactic CR contribution is made of protons only and the transition from galactic to extragalactic CRs is completed at $10^{18}$ eV. The transition occurs through a dip, produced by the onset of pair production, a feature which is very well defined in the CR spectrum. The chemical composition in the transition region changes suddenly \cite[]{dip4} from an iron dominated Galactic one to a proton dominated extragalactic one.

In both scenarios the Galactic CR spectrum is cut off well below the ankle. The lowest energy part of the extragalactic CR spectrum is however affected, in both scenarios, by unknowns related to the propagation of extragalactic CRs in weak magnetic fields that might be present in the intergalactic medium, therefore the contribution of extragalactic sources to CRs in the energy region around a few $\times 10^{7}$ GeV remains poorly constrained. This uncertainty makes the need for an additional class of sources in the transition region somewhat weak.

\section{Discussion and conclusions}
\label{sec:discuss}

The problem of explaining the origin of CRs as observed at the Earth is a complex one: it requires us to find the sources, describe the acceleration of nuclei (and in fact of electrons as well) in such sources, propagate all particles through the Galaxy taking into account diffusion and losses, and finally correct for local effects (such as the solar modulation at low energies). One can easily realize that this is a hard task, and yet one in which we have been successful in many respects, less in others. 

The most plausible sources of Galactic CRs remain SNRs, although a solid proof of the supernova paradigm satisfying all scientific standards has not been obtained as yet. The acceleration process is most likely diffusive shock acceleration in its non-linear version, which accounts for the dynamical reaction of accelerated particles on the shock and for the crucial phenomenon of self-generation of amplified magnetic fields in the acceleration region. In this respect we have been successful, in that several versions of the theory exist and the different formulations compare well with each other \citep[see for instance][]{comparison}. Our semi-analytical approach has the advantage, when compared to all others, of simplicity and reduced computation time, which is the very reason why it is increasingly more adopted in complex hydrodynamical codes of the SNR evolution including the reaction of CRs. 

In the present paper we completed the theoretical framework previously put forward by our group with the inclusion of accelerated ions, which we find to contribute considerably to the shock modification. 
We show that the spectra of different chemicals observed at the Earth can be reproduced reasonably well and we also calculate the implications of acceleration of nuclei on the gamma-ray emission of an individual SNR. The all-particle spectrum of CRs is also reproduced in a satisfactory way.

The one presented here is not the first attempt at describing acceleration of nuclear species in SNRs in the context of non-linear diffusive shock acceleration: \cite{berevolk} published previous work on the topic using a numerical approach to the acceleration problem. Unfortunately a detailed comparison of our results with their work is made difficult by the fact that the procedure they used to include nuclei and the amplification of magnetic field are not fully illustrated: actually it is not even clear to us whether the nuclei were introduced in the non-linear chain or rather included as test particles in a shock structure mainly modified by CR protons. The spectra obtained in that paper for the individual nuclear species were very flat (and in fact concave) leading the authors to require a diffusion coefficient with a strong energy dependence, $D(E)\propto E^{0.75}$. Such diffusion is however known to result in the breaking of the diffusive approximation well below the knee and consequently in excessive CR anisotropy, at odds with observations.

More details were given in the work recently presented by Ptuskin et al. \cite{pzs10}. Their calculation consists again in a finite differences scheme for the solution of the coupled transport and fluid equations, as for \cite{berevolk}. The authors discuss the effect of assuming a large velocity of the scattering centers, which leads to steepening of the resulting spectra (and to a reduction of the CR acceleration efficiency). As discussed in \S\ref{sec:spectra}, this very important issue, that was also discussed by Caprioli et al. \cite{crspec}, is very poorly known: the velocity of the scattering centers can hardly be estimated reliably, and even more important the helicity of the waves is unknown. In fact, if the helicity is not chosen properly (but arbitrarily) the net effect may well be that of flattening the spectra even further. In the paper by Ptuskin et al. \cite{pzs10}, as well as in ours, it is assumed that an estimate of the speed of the waves is represented by the Alfv\'en velocity evaluated in the amplified magnetic field at the shock. In \cite{pzs10} the authors assume that the waves downstream of the shock move in a direction opposite to that of the shock. In the present paper we make the more conservative assumption that somehow waves are isotropized downstream, so that $\langle v_{W}\rangle=0$. Both choices lead to spectra of accelerated particles which are steeper than the standard predictions of non-linear theory. In the absence of this very important ingredient it is not possible to explain CR spectra as observed at Earth. The main difference between our assumption and that made by Ptuskin et al. \cite{pzs10} is that in the latter case the spectral slope is strongly dependent on the Alfv\'enic Mach number, which leads the authors to suggest that some, yet unknown, self-regulation mechanism might be at work to guarantee the appropriate level of field amplification. 

\cite{pzs10} also made an attempt at taking into account the possibility that different classes of SNRs may contribute to the CR spectrum at Earth. However the relative abundances of chemicals injected in each class is the same and only the environmental conditions around the SNR (for instance gas density and temperature) are changed in the different cases. 

Although the introduction of these and other imaginable complications does not present any serious technical problem for the calculation procedure discussed here, we decided to focus on the main physical ingredients and avoid additional assumptions that, in our opinion, take away clarity from the results rather than making them more reliable. In our view, at the present stage of our knowledge, the assumptions that are required to carry out more sophisticated calculations are too many to lead to a scientifically trustable result. 

This leads us to a question that is of relevance for the investigation of the problem of CR origin from the theoretical point of view, namely {\it ``can we move beyond this point in a way that really adds to our knowledge of the problem?''}. 

There are clearly several open issues that are worth thinking about: (1) why are the observed spectra of protons and nuclei somewhat different? (2) what is the effect of the winds of the presupernova stars? (3) what is the best way to describe magnetic field amplification and/or complex topology of the magnetic field in the shock region? (4) how can we reliably describe the escape of CRs from a SNR? and (5) what is the Physics of injection of nuclei in a collisionless shock?

It is possible (although far from guaranteed) that each one of these questions may be tackled from first principles, and if this ever happens, then the challenge will be putting together the conclusions in a convincing way to reproduce the CR spectra observed at Earth. Taking into account the wide variety of situations that are realized in Nature and the fact that in principle the CR spectra observed at Earth are the result of the convolution of so many different environmental situations, even this latter task would be formidable. At present, we think it is difficult to envision going beyond the point already reached and still maintain the same standard of scientific credibility. On the other hand, the modeling of individual SNRs which leads to predict their appearance in different frequency ranges may provide several breakthroughs in the way SNRs accelerate nuclei and electrons, and this is probably the most promising avenue to follow for future progress.

\section*{Acknowledgments}
We are grateful to the referee, Don Ellison, for his valuable comments. 
This work was partially supported by ASI through contract ASI-INAF I/088/06/0.

\label{lastpage}

\end{document}